# Developing Centimeter-scale-cavity Arrays for Axion Dark Matter Detection in the 100 Micro-electron-volt Range

**Erik W. Lentz**[1] **Christian R. Boutan**[1] **Matthew S. Taubman**[1] **Kevin L. Gervais**[1]

[1]*Pacific Northwest National Laboratory, Richland, WA 99354 USA*

   *E-mail:* erik.lentz@pnnl.gov

Abstract: The cavity haloscope technique has been the most successful approach to date in searching for axion dark matter, owing to a confluence of factors at the GHz scale including the macroscopic size of the axion-to-photon converting cavity volume, the sophistication of present radio-frequency/microwave technologies including quantum amplifiers, and the location of the quantum limit temperature. These factors scale in a disadvantageous way overall as searches move up the axion mass/frequency scale, with the quantum limit noise temperature scaling linearly with frequency $T_{\rm SQL} \sim f$, the effective single cavity volume scaling as the inverse frequency cubed $CV \sim f^{-3}$, and the axion-coupled cavity mode quality factor shrinking as $Q \sim f^{-2/3}$ for copper cavities, necessitating the search for remedies. One approach is to make up the loss in volume using an array of efficiently packed matched cavities coordinated in space and time to act as a single axion-to-photon converting array. This paper presents PNNL's progress in developing technologies for cavity array axion haloscope in the $m_a \sim 100 \ \mu{\rm eVc}^{-2}$ mass range including the design of moderate scale cm-diameter cavities and their fabrication process using electric discharge machining, the development of mode tuning mechanisms towards a re-entrant style combination tuning rod and coupler, mode matching, and RF readout. The result is the first demonstration of a tunable array of matched cavities with axion-coupling modes in the $f_0 \in [22.88, 22.93]$ GHz ($94.62-94.83 \ \mu{\rm eVc}^{-2}$) range. Prospects for future larger arrays leading to viable axion DM searches of this type in this mass range are discussed.

# Contents

# 1 Introduction

The 'invisible' axion has been a tantalizing candidate for solving both the strong-CP problem of particle physics [1–3] and the dark matter problem of astrophysics and cosmology [4–6] for nearly five decades, but only in the last decade has there been significant progress in its search [7–63] to the stringent benchmark models of KSVZ (Kim-Shifman-Vainshtein-Zakharov) model [64, 65] and DFSZ (Dine-Fischler-Srednicki-Zhitnitsky) [66, 67]. The next decade is shaping up to be increasingly intense, with the list of proposed and operating searches growing rapidly [68, 69], which is necessary given the poorly predicted axion mass.

That being said, only one technique has thus far been able to reach the KSVZ and DFSZ limits in the micro-to-milli-eV mass range: the cavity haloscope (HS) [70]. The cavity HS consists of a cold microwave cavity threaded by a static magnetic field and couples to the ambient axions via an inverse-Primakoff process rooted in the classical interaction Lagrange density term $\mathcal{L}_{a\gamma\gamma} = -g_{a\gamma\gamma}\varphi \mathbf{E} \cdot \mathbf{B}$ where $g_{a\gamma\gamma}$ is the axion-electromagnetic coupling strength, $\mathbf{E}$ is the electric field, and $\mathbf{B}$ the magnetic field.

Axion DM is considered to be cold and highly degenerate [71] and therefore able to uniformly and coherently drive a high-quality mode in the cavity with net axion coupling as quantified by the form factor

$$C_{\text{mode}} = \frac{\left(\int_V d^3x \mathbf{E}_{\text{mode}} \cdot \mathbf{B}\right)^2}{\left(\int_V d^3x |\mathbf{E}_{\text{mode}}|^2\right)\left(\int_V d^3x |\mathbf{B}|^2\right)}, \qquad (1.1)$$

where $V$ is the region occupied by the mode and $\vec{B}$ is the net magnetic field. The expected power deposited into a single cavity haloscope from a monotone source and averaged over times much longer than the coherence time of the cavity mode is found to be approximately

$$\langle P_{\text{cav}}\rangle(f) \approx \frac{\epsilon_0 \alpha^2 c^2}{\pi^2 f_a^2} g_\gamma^2 f \left\langle |a|^2 \right\rangle (f) V |B|_{\text{max}}^2 C Q_L T_{f_0}(f)$$

$$\approx 1.06 \times 10^{-26} W \left(\frac{g_\gamma}{0.36}\right)^2 \left(\frac{\rho}{0.45 \text{ GeV/cm}^3}\right) \left(\frac{V}{0.01 \text{ L}}\right) \left(\frac{B_{\text{max}}}{10 \text{ T}}\right)^2 \left(\frac{C_{\text{mode}}}{0.5}\right) \left(\frac{f}{10 \text{ GHz}}\right) \left(\frac{Q_L}{10^4}\right) T_{f_0}(f).$$

Here $g_\gamma$ is the unitless coupling strength, $f_a$ is the Peccei-Quinn symmetry breaking energy scale, $\rho_a$ is the local axion DM density, $\epsilon_0$ is the permittivity of free space, $c$ is the speed of light in vacuum, $\alpha$ is the fine structure constant, $V$ is the cavity volume, $B_{\text{avg}}$ is the average value of the applied magnetic field, $Q_L$ is the loaded quality factor of the target cavity mode with envelope shape $T_{f_0}(f)$ expected to be of Lorentzian form. Tuning of the axion-coupled mode's frequency to scan a broad range of potential axion masses is most often accomplished by alteration of the mode geometry using one or multiple tuning rods run through the cavity.

Single-cavity experiments have proven difficult to reach DFSZ sensitivity and broadly scan in a timely manner at frequencies beyond 1.5 GHz. The already meager power of Eqn. 1 is further confounded above 20 GHz by the poor scaling relation of the effective cavity mode volume with inverse cubed cavity mode frequency when presumed to track alongside the Compton volume $V_C = \lambda_C^3 = (h/m_a c)^3 \sim f^{-3}$. Further, the mode's uncoupled quality factor scaling $Q \sim f^{-2/3}$ due to skin-depth current losses, presuming the use of normally conducting materials such as copper. These factors result in rapidly decreasing signal power as frequency increases with $P_{\text{cav}} \sim f^{-8/3}$, which implies a scaling in the fiducial signal-to-noise ratio for a HS at the standard quantum limit (SQL) of $SNR \sim f^{-14/3}$. See the single-cavity estimates of Figure 1 (orange line) for the impact on sensitivity to the axion-electromagnetic coupling strength $g_{a\gamma\gamma}$.

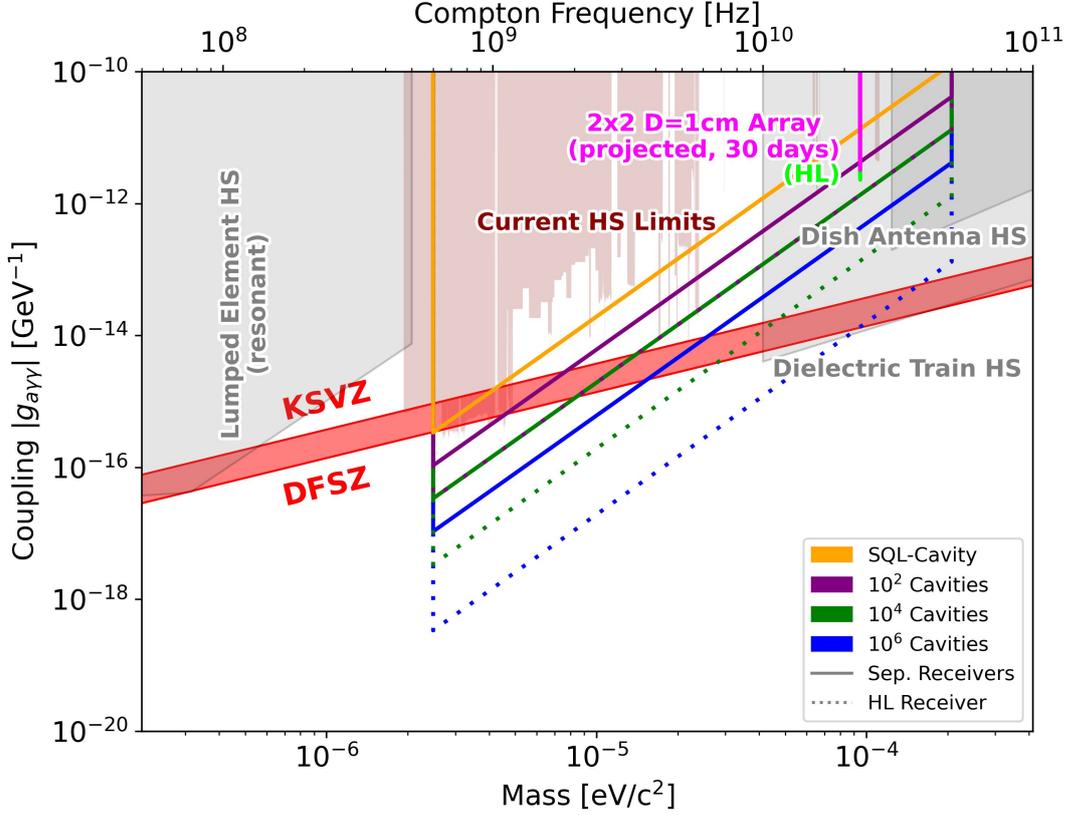

**Figure 1**: Haloscope sensitivity estimates for single cavities (orange), coherent cavity arrays of size $N_{cav} = 10^2, 10^4, 10^6$ (solid lines), and HL cavity arrays (dashed lines). Existing haloscope limits are in dark red in the background. A cavity has Compton volume $V_{cav} = \lambda_C^3$ and $Q_L = 10,000 (10\ GHz/f)^{2/3}$. A near perfect packing fraction is assumed for the arrays to keep them within a coherent patch. Projections for different haloscope techniques in adjacent bands are displayed in gray. Lumped Element HS (DMRadio) is given $Q = 100,000$ and volume 500 liters below 80 MHz and the Compton volume $V = \lambda_C^3$ above [72]. Dish Antenna HS (BREAD) has 10 m$^2$ collecting area [73]. Dielectric Train HS (MADMAX) has 80 disks of 1 m$^2$ area and boost factor taken from [74]. All projections assume $B = 10$ Tesla, SQL readout noise scaling, and 10 years total observation time unless otherwise stated. The projected sensitivity of the 2x2 cavity array demonstrated in Section 4 over the frequency range [22.88,22.93] GHz and 30 days observation time is shown in magenta assuming separate receiver detection and in lime assuming a HL receiver.

This un-advantageous scaling has resulted in multiple next-generation axion HS experiments turning to arrays of matching cavities bundled together to take advantage of large bore (O(1 meter)) moderate field (O(10 Tesla)) magnets [68, 69, 75]. Well-packed cavity arrays with extent much smaller than the de Broglie length of a virialized Milky Way DM halo [76] $\lambda_{dB} \gtrsim 1000$ meters $\times$ ($\mu$eVc$^{-2}$/m$_{DM}$) $\times$ (300 km s$^{-1}/\sqrt{\Delta v_{DM}^2}$)$^2$ may use the coherence of a classical axion field to bring an advantage of $SNR \sim \sqrt{N_{cav}}$ when the measurement uncertainties due to random noise of each cavity are independent, see the solid lines in Figure 1. This advantage may be increased up to $SNR_{HL} \sim N_{cav}$ in the Heisenberg Limit (HL) where the uncertainty of the array readout responds like a single sensor [77], see the dashed lines in Figure 1.

The capabilities to fabricate and bundle cavities using copper and ceramic surfaces and tuning using mechanical rods are available and practiced up to the 10 GHz ($\lambda_C \approx 3$ cm) scale and are scheduled to be implemented by various experimental collaborations in the next decade. Looking beyond 10 GHz, it is less clear what technique(s) will lead in the pursuit to directly detect axion DM [68, 69]. See Figure 1 for sensitivity projections of two leading concepts, the Dish Antenna HS [73] and Dielectric Train HS [74], both of which rely on yet-to-be-proven technologies that have been in development for more than a decade [78, 79]. We take the perspective here that it is prudent to simultaneously pursue variations on the proven cavity HS approach, exploring what steps can be taken to maintain it as a practical technology well beyond 10 GHz despite scaling disadvantages.

This article presents study to assess technology readiness for scalable cavity array HSs in the 100 $\mu$eVc$^{-2}$ axion mass range. Section 2 provides the authors' requirements for an effective operational cavity array haloscope and are used as a guide to orient an R&D study. Section 3 assesses to what measure these requirements may be fulfilled by testing the basic design, engineering, and fabrication capabilities needed to build such HSs, resulting in a small-scale tunable cavity array. Section 4 presents a demonstration of synchronous tuning operation of a prototype 2x2 cavity array over the frequency range $f_0 \in [22.87, 22.92]$ GHz (94.58 − 94.79 $\mu$eVc$^{-2}$). Section 5 summarizes this work's progress, outlining how the remaining challenges to 100 $\mu$eVc$^{-2}$ cavity array HSs with DFSZ sensitivity may be addressed.

## 2 Haloscope Requirements and Study Goals

This section lists high-level requirements for what makes a successful cavity haloscope, followed by this study's goals in determining viability for a 100 $\mu$eVc$^{-2}$ cavity array. This is not intended as a complete or singular list, but reflects the primary drivers common to modern cavity haloscopes.

- **Maximize coherent axion signal power**: The read out power spectrum from converted axions in a haloscope using a single cavity mode imbued with an externally sourced magnetic field is proportional to $\langle P_{cav} \rangle (f) \propto fVB^2C_mQ_LT_{f_m}$, generalizing to a matched array of coherently combined cavities as the sum of power outputs $\langle P_{cavs} \rangle (f) \propto \sum_i B_i^2 V_i C_i Q_i \beta_i T_{f_i}$. A cavity array haloscope should seek to maximize the total axion power output.

    > *Goal*: This work seeks to show that cavities with a high-quality axion-coupled mode in the 100 $\mu$eVc$^{-2}$ mass range can be consistently produced in well-packed arrays such that the array of mode outputs can be coherently combined.

- **Broaden frequency range**: The unknown axion mass is taken as warrant to cover a wide frequency/mass range. Scanning high-quality/narrow-bandwidth modes by changing the geometry of the cavity is the most widely used approach to fulfill that commission.

    > *Goal*: This work seeks to implement cavity tuning mechanisms able to be applied across the cavity arrays while maintaining high conversion power.

- **Rapidly tuning array**: To work as a coherent unit, the chosen mode's output from each cavity must be aligned rapidly and to high precision across the array in mode center frequency, Q-width, coupling, and path length such that their outputs are synchronous and may be directly co-added. The complexity of this matching action scales quickly with the number of degrees in the system and their interactions.

    > *Goal*: This work seeks to align the fabricated cavity arrays about prescribed mode parameters in the tuning range such that the cavities can be coherently combined.

- **Minimize readout noise**: The sensitivity of a HS is ultimately determined by the comparison of signal strength to measurement uncertainty (noise). Haloscopes achieve this in several ways, including lowering their physical thermal temperature as much as is feasible and reducing introduced noise in the readout process. The increasing complexity of achieving this uniformly across a cavity array's readout is a standing challenge among HSs.

    > *Goal*: This work seeks to directly combine cavity outputs as a means to pursue HL readout methods.

## 3 Haloscope Array Hardware Development

This section documents our assessment of technologies for fabricating fundamental HS hardware: cavities, RF mode tuners & antenna couplers, and RF read-through. The most practical and compatible technologies at the time of the study were chosen to construct a preliminary small-scale haloscope array, to be demonstrated in Section 4.

We start the description of the design consideration at the cavity array. In order to arrive at a cm-scale cavity design, the authors chose to begin with a rudimentary exploration of machining 10 mm and 5 mm diameter holes in a 6.35 mm (1/4 inch) blank of (Residual Resistive Ratio) RRR-30 (C101 alloy) copper, chosen for the significant gain in RF cavity Q for potential cryogenic tests. The geometry for these cavities vacated from the blank is shown in Figure 2. To close the vertical electrical circuit of the $TM_{010}$ mode for each cavity are two end plates made of the same RRR-30 copper blanks pressed firmly onto the center cavity bore plate. A thick gold plating is to be applied to the contact surfaces to ensure robust conductive continuity. Knife-edge contact was also considered but deemed to impractical to implement consistently across an extensive plate with many cavities.

To get a sense for the tolerances needed to produce matched cavities in the above right cylindrical geometry, consider the TM and TE modes with center frequencies given by

$$f_{mnp}^{TM(TE)} = \frac{1}{2\pi\sqrt{\mu\varepsilon}} \cdot \sqrt{\frac{x_{mn}^{(\prime)2}}{R^2} + \frac{p^2\pi^2}{d^2}}, \qquad (3.1)$$

where $x_{mn}$ are the nth roots of the Bessel function $J_m(x)$, and $x'_{mn}$ are the n-th roots of the derivative of that Bessel function, $J'_m(x)$. The frequency of the $TM_{010}$ is thus $f_{010}^{TM} = \frac{1}{2\pi\sqrt{\mu\varepsilon}} \cdot \frac{x_{01}}{R} = \frac{114.75\,[\text{MHz}\cdot\text{m}]}{R\,[\text{m}]}$, where $x_{01} = 2.405$, and we have taken $\mu = \mu_0$ and $\varepsilon = \varepsilon_0$ since the cavity is non-magnetic and air-filled. Table 1 shows this frequency modeled for both $R = 5$ mm ($D = 1$ cm) and $R = 2.5$ mm ($D = 5$ mm) sized cavities, as well as those of their first three TE modes. The TE modes are located at far higher frequencies than the desired $TM_{010}$ mode due to the relatively shallow height of the cavity, thus avoiding mode crossings. Generally, TE modes can and should be selected against through correct coupling techniques.

Even for a perfectly polished and formed right cylindrical cavity, the maximum possible unloaded Q of the $TM_{010}$ mode is limited by skin depth $\delta$ of the material.[1] The quality factor may be estimated by $Q = \frac{d/\delta}{1+d/R}$ where $d$ is the cavity height, and the skin depth is given by $\delta = \sqrt{\frac{\rho}{\pi f \mu}} \times G_{HF}$ where $\rho$ is the resistivity of the material and $G_{HF}$ is a factor that tends to unity far away from material resonances [80], which is the case for the copper and gold used here. This gives 304.5 nm, at room temperature for copper at $f = 45.9$ GHz ($D = 5$ mm), and 430.6 nm at $f = 22.95$ GHz ($D = 1$ cm). This in turn gives the room temperature Q values shown in Table 1. Clearly, the small dimensions and mode dependence on skin effect deeply limit the Q at room temperature, much more severely than for big RF cavities currently used in haloscopes at the GHz scale.

As temperature is reduced to cryogenic levels, the preferred environment for a haloscope DM search, resistivity drops significantly for most metals and their alloys [81]. For our case, copper RRR-30, $\rho = 1.68 \times 10^{-8}\,\Omega \cdot m$ at room temperature. Below about 20°K, $\rho$ levels out at around $5.3 \times 10^{-10}\,\Omega \cdot m$ for the same material. This corresponds to skin depths of 54.08 nm for $f = 45.9$ GHz, and 76.48 nm for $f = 22.95$ GHz. The RRR-30 copper underpins the design strategy of these prototype arrays, supporting the bulk conduction of electrical currents running vertically in the cavity sidewall for the $TM_{010}$ mode.

---

[1]Note that certain TE modes are not limited in this way, and thus exhibit much, much higher Q values. However, to date no one has figured out how to use them to detect axions as the E-field isn't linear.

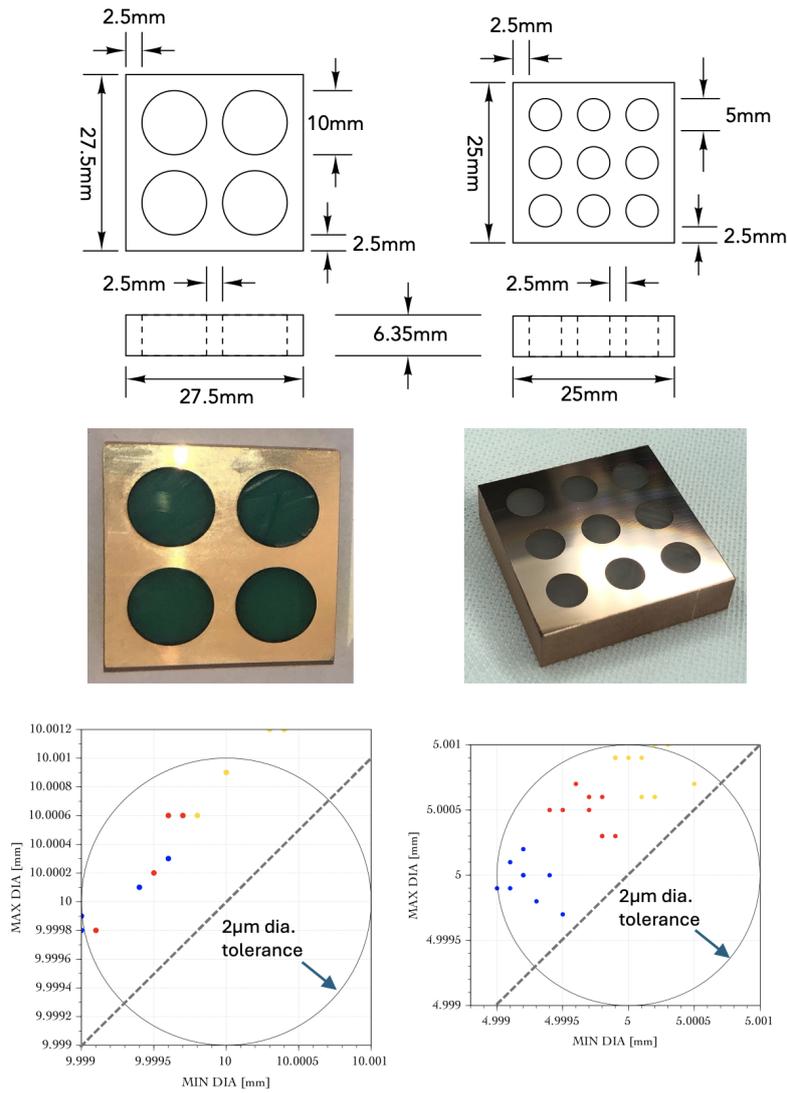

**Figure 2**: (Left column) Prototype 2x2 cavity array. (Top) Draft of the cavity center plate. (Middle) Photo of example cavities after EDM by iMC Intertech (16µin Ra, or 0.4µm Ra), silicone plugging, diamond turning finishing by Surface Finishes, and gold plating (32 µin Ra, or 0.8µm in Ra) by ProPlate. (Bottom) Radial tolerances measured from each cavity from three example machined plates, color coordinated to each plate. (Right column) Same sub-figures for the 3x3 array, but with phto at middle taken prior to the gold plating.

The precision to which the cavities must be fabricated to satisfy an uncertainty in the mode

| $R$ | $f_{010}^{\text{TM}}$ | $\Delta f_{010}^{\text{TM}}$ | $\delta f_{010}^{\text{TM}}$ | $Q$ (RT) | $Q$ (Cryo) | $f_{111}^{\text{TE}}$ | $f_{011}^{\text{TE}}$ | $f_{112}^{\text{TE}}$ |
|---|---|---|---|---|---|---|---|---|
| 5 mm | 22.95 GHz | 23 MHz | 3.5 MHz | 6,496 | 36,570 | 29.43 GHz | 43.52 GHz | 50.37 GHz |
| 2.5 mm | 45.90 GHz | 93 MHz | 7.8 MHz | 5,891 | 33,170 | 42.33 GHz | 76.85 GHz | 58.85 GHz |

**Table 1**: Properties of desired TM010 mode and frequencies of nearby TE modes are shown for two different radii cavities. Cavity length is 6.35 mm. Room temperature (RT) and in the cold limit (Cryo) $Q$ are given. Uncertainty in mode center frequency, $\Delta f$, are based on a 5-micron machining uncertainty in the radius. Line widths $\delta f$ are estimated at room temperature. $Q$ values, errors and line widths are discussed directly.

center frequency can be approximated as

$$\Delta f_{010}^{\text{TM}} \approx f_{010}^{\text{TM}} \times \frac{\Delta}{R}. \tag{3.2}$$

For $\Delta \approx 2 \times 10^{-4}$ inch $\approx 5.08\,\mu\text{m}$, a precision one may reasonably expect from commonly found traditional milling and ream-finishing techniques, this gives a $\Delta f \sim 23\,\text{MHz}$ uncertainty in mode frequency for the 10 mm diameter cavities and 93 MHz uncertainty for the 5 mm diameter cavities, as shown in Table 1. This compares to the line widths of the two cavities $\delta f_{010}^{\text{TM}} = f_{010}^{\text{TM}}/Q$ that at room temperature ideally come to be 3.5 MHz in mode frequency for the 10 mm diameter cavities and 7.8 MHz for the 5 mm diameter cavities, values that are collected in Table 1. The mode width is to be used as the cavity precision benchmark to give the cavities in an array well-matched tuning maps. Traditional machining techniques are therefore too coarse by an order of magnitude compared to the ideal room temperature mode width benchmark. Secondly, this factor is bigger for the smaller cavity than the larger. This indicates that smaller cavities are in general going to be more demanding to fabricate accurately to a mode-width standard than larger ones, as expected.

An alternative high-precision machining techniques is EDM (electrical discharge machining). To fulfill the requirement that the uncertainty in the mode center frequency be approximately that mode's ideal line width at room temperature for the 5 mm diameter cavities, those cavities would need to be machined to a radial precision of $\Delta = \frac{R \cdot \delta f}{f_{010}^{\text{TM}}} = \frac{2.5 \times 10^{-3}\,\text{m} \cdot 7.79\,\text{MHz}}{45.9\,\text{GHz}} = 424.3\,\text{nm}$ from round. This in turn leads us to the general diameter tolerance for EDM cavity fabrication of 1 μm used for both the 10 mm diameter and 5 mm diameter cavity arrays.

The cavity machining and coating was performed in stages using several company partners. The machining of the cylindrical voids was performed by iMC Intertech, who verified that they were able to cut to within a 1 μm tolerance in most cases. See Fig. 2 for the measured cavity precision in three completed 10 mm diameter arrays and 5 mm diameter arrays. Copper oxidizes quickly, therefore as part of the machining instructions we required the machinists to immediately wrap the completed parts with Coppertex® paper. This paper protects copper from corrosion and tarnish. The forces involved in producing the cylindrical voids in the cavity plate produced irregular bulging and deformation at the micron scale on the top and bottom surfaces, which could have interfered with the electrical connection between the cavity side walls and the end plates. To mitigate this, the voids were plugged with silicone and then diamond turned machined at Surface Finishes to a tolerance of 16μin (0.4μm) Ra to produce a flat surface. The silicone plugs enable precision gold plating up to the edge of the cylinder and kept the machining edge sharp. One cannot directly coat copper with gold and expect it to be robust. An intermediary must be used. Further, this intermediary was required to be non-magnetic to avoid unwanted stress cycling when used in an axion search, also that the intermediary not become superconducting at low temperatures. For these reasons common intermediaries such as nickel were ruled out. Rhodium was selected after some consideration. Though it does become superconducting at 325 μK, this is unlikely to be a problem for foreseeable axion detectors operating above 20 GHz where the SQL temperature is orders of magnitude higher. The resulting blanks were then plated at ProPlate first with a 5 μin (0.13μm) layer of rhodium, then with a minimum 20 μin (0.51μm) layer of "hard" gold (per Mil-G-45204). Pictures of example completed 10 mm diameter and 5 mm diameter cavity arrays are shown in Fig. 2.

Tuning a right cylindrical cavity has been pursued in multiple ways in the literature including via right cylindrical tuning rod(s) running near the length of the cavity and mechanically moved within the bare cavity volume [7], and using cylindrical post(s) inserted from an end plate in the re-entrant style [82]. A re-entrant style with a simple post was chosen for its simplicity and repeatability across the cavity array, further for its substantial tuning range with acceptable impact to mode width, see Fig. 3.

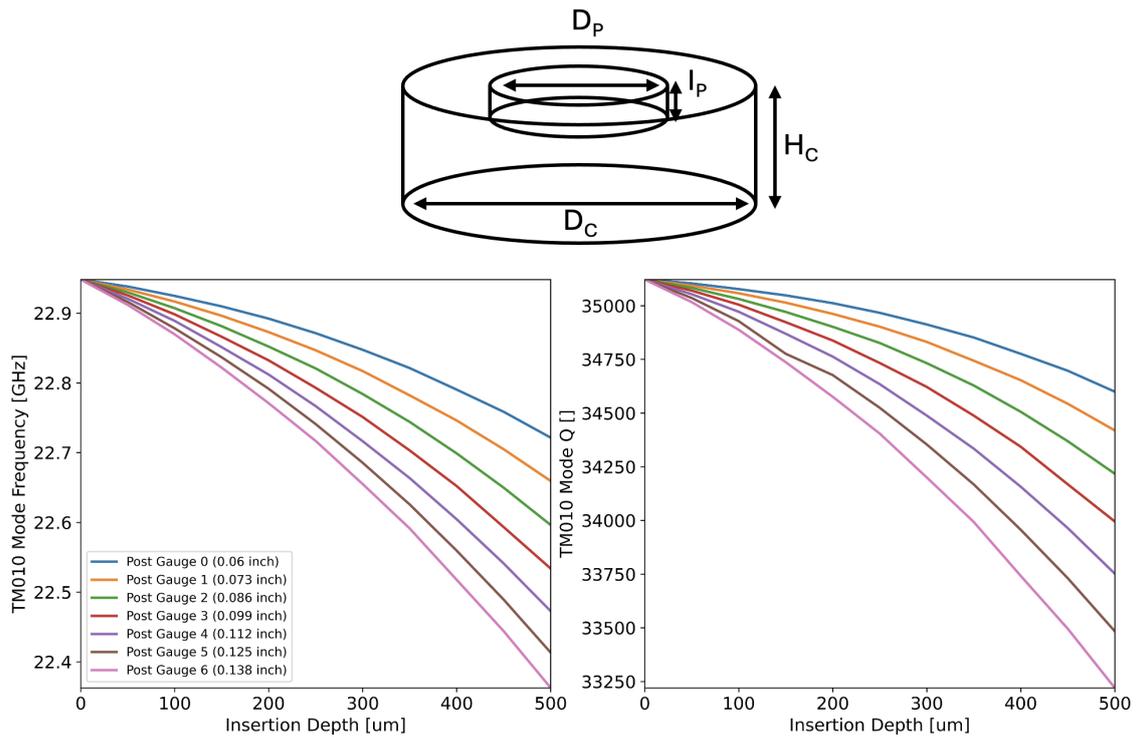

**Figure 3**: (Top) Illustration of cavity model with re-entrant post inserted. (Bottom) COMSOL simulation results for RRR-30 copper cavity of diameter $D_C$ = 10 mm, height $H_C$ = 6.35 mm held at 1 Kelvin. (Left) TM010-like mode center frequency tuning with post insertion depth. (Right) TM010-like mode $Q$ versus post insertion depth.

Coupling techniques for injecting RF signals into and extracting them from the cavity mode are constrained to have form factor small enough to fit inside the inter-cavity spacing. After some consideration, wire antennas were chosen as the couplers, inserted from the top and bottom end plates and concentric with the cylinder and re-entrant post. In fact, the post itself is made from the antenna co-axial cable stock (Pasternack's PE-034SR stock, outer conductor diameter 0.034 inch (0.86 mm), inner conductor diameter 0.008 inch (0.2 mm), and which is held electrically independent from the cavity end plate through which it emerges. This choice constrains each cavities' tuning and coupling mechanisms to act together. The post-antenna is labeled as the "strong port" and is prepared with outer conductor and dielectric layers stripped to expose 3/16 inch (4.76 mm) inner conductor. At the opposite end plate is the "weak port", also made of the same coaxial cable stock, but with only the center conductor exposed for a length of 1/16 inch (1.59 mm) and placed so as to be nearly uncoupled from the cavity mode. The tuning post and antenna coupling configuration can be seen in Fig. 4. Feed through holes in the end plates were machined to a diameter of 1 mm.

Mechanically inserting and extracting the coaxial cables from each cavity must be precise and repeatable, such that the output signals from each cavity may be combined coherently across the modes' mean width. This requires the near-matching of mode center frequencies, Q-widths, couplings, and overall phase length, while having only three tunable parameters to each cavity (including a phase-shifter), though recall the degeneracy between the re-entrant tuning port and the strong port antenna. This requirement conferred to tolerances on the center frequency tuning of 5% of mode width, mode width tuning to 0.05% of mean mode width, coupling strength of 0.1 dB, and global phase length to within 10 degrees. Automated tuning was considered using commercial piezo-electric with translational motion, however the form factor for such motors was considered too large and the projected heat load from operating an array of motors at scale too high. Instead a system of threaded screw adjusters translating rotational to translational motion was chosen, see Fig. 4. This class of tuning mechanisms was chosen as it lent itself better to more petite form factors and has precedent for cryogenic testing, with rotary motors located at room temperature where heat load is less of a concern and and motion may be transmitted by rotary feed through via a flexible rotary shaft into vacuum and down the temperature stages to the cavities.

The analogous tuning mechanism implemented for this study's warm testing was achieved by using elaborate external threaded screw adjustments in a jig, see Fig. 4. The screw pitch gave an insertion/extraction rate of +/- 0.05 inches (1.27 mm) per-rotation, allowing for precision motion of the antennae of 140 µin (3.5µm) per-degree which was found to be more than sufficient for room temperature tests. A spring loaded design was used to remove the slack in the screw threads and increase the concentrical tolerance. The design had a pointed set screw interfacing to a dimpled surface. This interface created a low-friction contact point under spring compression to improve turning precision, also freeing the coax cable from twisting when the threaded screw was turned. Part of the challenge to this mechanical mechanism was the non-linear forces created by the coax cable leads. The spring force in the mechanical assembly required enough force to overcome the coax cable tension. However, too much spring force created excessive friction which would cause the coax cable to spin.

The readout of RF transmission measures in open air for the individual cavities and the combined array was performed via vector network analyzer (VNA, Keysight FieldFox N9952B) and directed through a network of two-way switches (using Mini-Circuits RC-8SPDT-A18 switch banks), indicated on the schematic of Fig. 5. Inputs and outputs could either be split amongst the individual cavities for the purpose of mode identification and tuning or combined to test the coherence of the array. The 1:4 channel dividers/combiners (Mini-Circuits ZC4PD-02263-S+) were used to coherently split the VNA input signal between the cavity channels and then combine the in-phase components of the outputs. Phase shifters could be added prior and/or after the cavities, but were left out of this schematic due to available phase shifters having insufficient range as discussed further in the next section. Background reduction beyond the natural shielding from co-axial cables and bulk materials surrounding the cavities was not pursued for the demonstration. Identical length cables and connectors were used to preserve constructive interference for the combined cavity sum measurement.

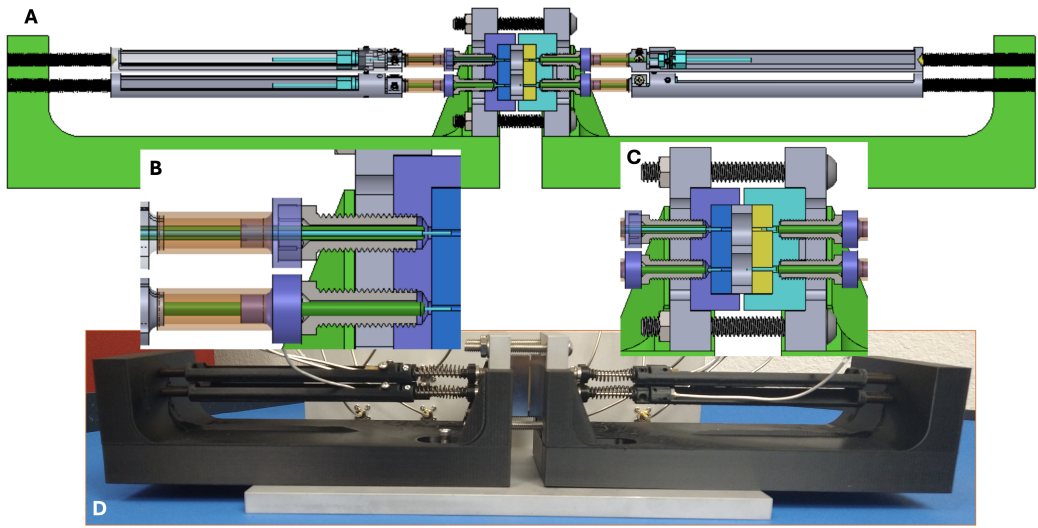

**Figure 4**: (A) Illustration of the mechanical tuning/coupling assembly with partial cutaways. (B) Close-up view of cavity plate, end plates with antenna feedthroughs, and outer enclosure and clamp. (C) Close-up view of feedthrough sub-assembly. (D) Photograph of the completed assembly anchored to a bulkhead for attaching cavity input/output RF coax cables.

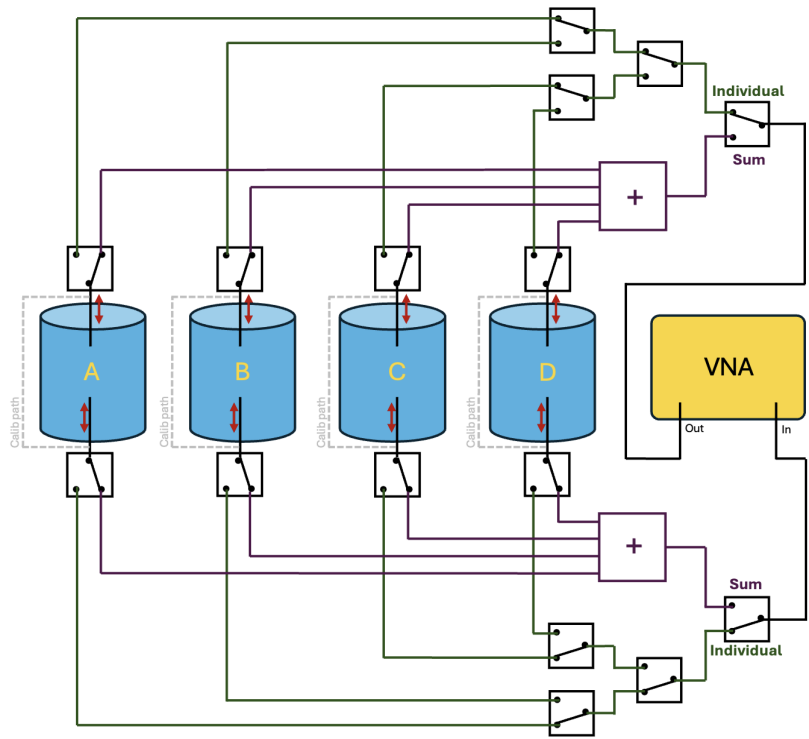

**Figure 5**: RF schematic for VNA readout of individual cavities and over the cavity array.

## 4 Matched Tuning Demonstration of 2x2 Array

Each cavity in the tunable 2x2 array described in the previous section may be individually tuned and its transmissions characterized by VNA. To demonstrate operation as a tunable matched array, where the transmissions from each cavity's TM010-like mode can be coherently combined, center frequency benchmarks were selected at 10 MHz intervals across the mode's observed tuning range of 22.88-22.93 GHz. The observed tuning range is somewhat narrower than that suggested by the simulations of Fig. 3 in part due to the degenerate coupling and tuning mechanism that were not a feature of the design simulations. Outside of this frequency range the strong port's degenerate coupling and tuning mechanism push the mode outside of the ideal of a high-quality critically coupled resonance, with the low frequency end having an over-coupled antenna with poor quality factor and the high frequency end having a higher quality factor but weakly coupled antenna.

Tuning of the cavities used an iterative cycling method, tuning each cavity in turn to a prescribed frequency/Q/coupling until converged to within tolerances. Three passes was found to be sufficient for convergence, though the amount of cross-talk between cavity tunings is limited in a 2x2 array. Even if the fine tuning procedure scaled as the number of cavities, which it is not likely given the increased number of nearest neighbor and next-nearest neighbor cavities in moderate sized arrays, this would be too slow for a DFSZ sensitive array containing thousands of cavities, eventually taking longer to tune than the integration time needed for the configured array to reach DFSZ sensitivity. Optimization of the cavity tuning procedure was explored, though a scalable solution has yet to be demonstrated.

Measurements of example S21 swept transmissions consecutively taken for each cavity in the array, for each 10 MHz-spaced mode tuning position in the above range, can be seen in Fig. 6. The observed mode quality factors were lower than expected by simulation, ranging from 1,000-3,000 from the bottom to top of the tuning range, due to additional introduced losses including the strong port coupling, surface tarnishing due to long-term air exposure, and potential surface scratches on the cavity end plates from antenna insertion errors. Further, variance in device path lengths were found to cumulatively form global phase errors between modes, preventing analogue coherent combining of the four cavity transmissions in the absence of in-line phase shifters with large range (+/- 100 degrees) as has already been mentioned. Collection of individual transmissions, removal of global phase errors, and digital combining of outputs show coherence across the array as can be seen in the overlaid gray lines in Fig. 6. The projected sensitivity of the demonstrated array scanning over the above 50 MHz range with 30 days of observation time using a 10 Tesla magnetic field can be seen in Fig. 1 where the magenta line assumes a digital combining of cavities with SQL noise levels, and the lime line assumes HL noise across the array.

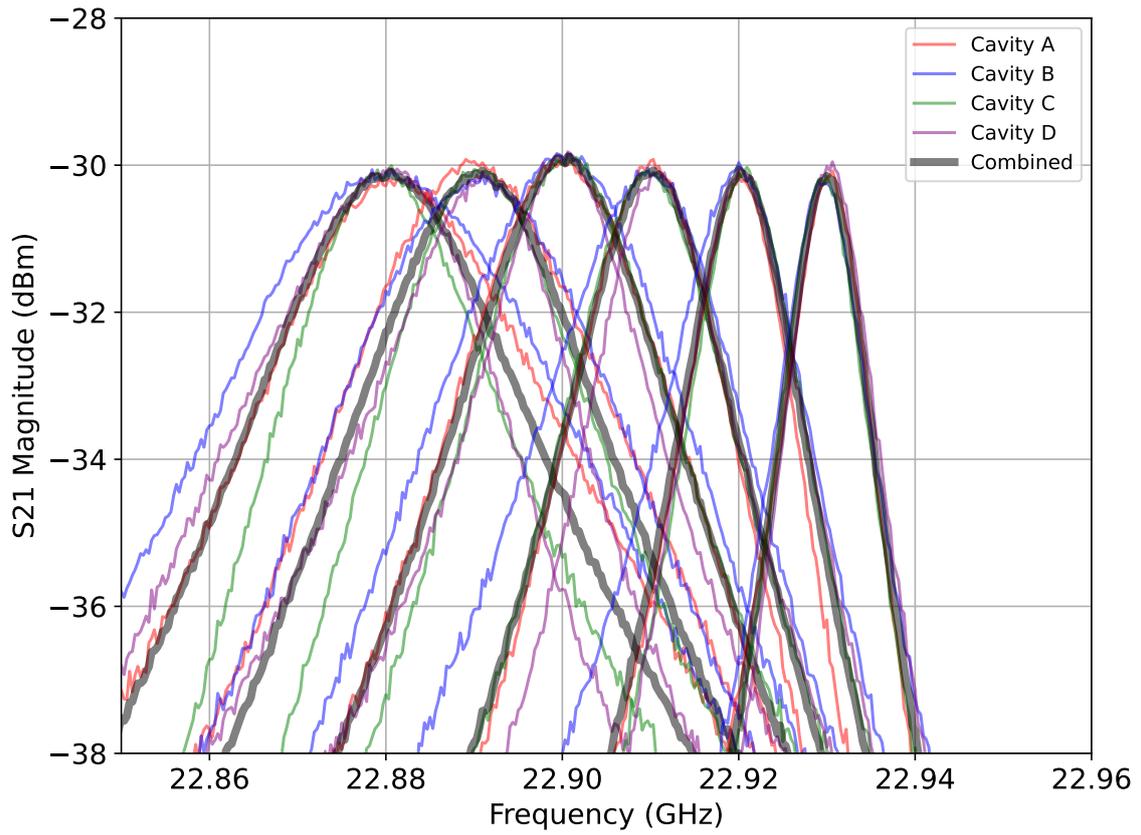

**Figure 6**: Individual and combined S21 transmissions of TM010-like modes from the 2x2 cavity array over the tuning range $f_0 \in [22.88, 22.93]$ GHz. Individual swept cavity transmissions are in color, while the digitally combined transmission is overlaid in gray.

## 5 Summary

This article has provided an account of efforts to develop cavity array axion haloscopes in the $100\,\mu\text{eVc}^{-2}$ mass range. Multiple goals were identified as crucial to produce a successful cavity array haloscope, including fabrication of high-precision tightly packed cavity arrays, the implementation of tuning mechanisms able to scan a chosen axion-coupled mode within each cavity over a broad frequency range, the ability to rapidly tune those modes to match a prescribed configuration such that their outputs may be coherently combined, and to introduce minimal noise during mode readout.

A successful means of fabricating cavity arrays to micron precision across the array was developed in this work. Prototype arrays were cut from a plate of RRR-30 copper using wire EDM in a regular lattice, with top and bottom contact surfaces diamond turned finished then gold plated using a rhodium binding layer, with endcap plates also finished and gold plated in the same manner. In an ideal use of this design, plates may be stacked and the lattice organized to fit the geometry of a magnet bore.

Replicable tuning and coupling mechanisms internal to each cavity in the lattice were then explored on one of the 2x2 arrays, ultimately settling on a pair of insertion mechanisms tied to the weak and strong ports, one on each endcap. Space limitations and complexity dictated that the rod insertion mechanism and strong port coupling be one and the same, creating a tuning-coupling degeneracy and limiting the effective tuning range as a sacrifice to simplicity. Control of the tuning mechanisms was achieved using an external threaded screw adjustments mounted on a jig.

Improving the technology up to the needed cavity count will require significant new developments in tuning miniaturization and fast fine tuning/alignment across the array, and high sensitivity readout. The technology to make centimeter-scale cavities exist, but needs a roadmap to demonstration for much larger arrays, in the hundreds to thousands on the way to sufficient scale to reach DFSZ sensitivity. The tuning mechanism exterior to the cavity must be miniaturized and automated. Insertion tuning requires large motors and/or jigs, taking up valuable space if placed amongst the cavities, lowering the cavity filling fraction in a magnet's bore space. Bore space may be saved and heat load reduced by displacing motors to outside the cold space with motion transmitted by rotary feed through via a flexible rotary shaft. Fuzzy clustering may provide a better approach to fast tuning.

## Acknowledgements

Pacific Northwest National Laboratory (PNNL) is operated by Battelle Memorial Institute for the United States Department of Energy (DOE) under Contract no. DE-AC05-76RL01830. This study was supported by the DOE Office of High Energy Physics Advanced Technology R&D subprogram. PNNL information release number PNNL-SA-219800.